# Ontology of Software Maintenance

Luis Olsina
*School of Engineering, UNLPam*
*GIDIS_Web*
General Pico, La Pampa, Argentina
olsinal@ing.unlpam.edu.ar

Pablo Becker
*School of Engineering, UNLPam*
*GIDIS_Web*
General Pico, La Pampa, Argentina
beckerp@ing.unlpam.edu.ar

María Fernanda Papa
*School of Engineering, UNLPam*
*GIDIS_Web*
General Pico, La Pampa, Argentina
pmfer@ing.unlpam.edu.ar

***Abstract*—Software maintenance is often a long and costly phase of the software lifecycle, encompassing activities such as defect correction, adaptation, code improvement, and functionality enhancement. Conceptual inconsistencies and terminological variation continue to hinder effective communication among researchers and practitioners. To address this gap, this paper presents the Software Maintenance Top-Domain Ontology (SwMaintTDO), a high-domain ontology developed within a five-tier ontological architecture. This ontology was constructed using competency questions, ontological analysis of documented maintenance processes, artifacts, and resources, and alignment with established software engineering standards. SwMaintTDO provides a unified and extensible conceptual model for representing key maintenance processes, roles, resources, artifacts, goal purposes, and the transition of target modifiable and context entities from current to updated maintenance situations. The applicability and value of SwMaintTDO are demonstrated through its application —via its concepts and relationships— to the process specification of a corrective maintenance strategy.**

## I. INTRODUCTION

David Parnas [1] observed in the early 1990s that "*Like human aging, software aging is inevitable, but like human aging, there are things that we can do to slow down the process and, sometimes, even reverse its effects.*" Software maintenance (SM) is one of the longest, most complex, and most expensive phases of the software lifecycle, often accounting for up to 90% of total lifecycle expenditure. This phase encompasses a broad spectrum of maintenance activities, including defect correction, adaptation, code improvement, and functionality/quality enhancement, as well as various types of maintenance artifacts, resources, roles, and goal purposes. Consequently, the variety in the large number of concepts in this field has generated conceptual complexity and terminological heterogeneity, both syntactic and semantic, hindering consistent understanding, effective communication, and appropriate conceptual alignment among researchers and practitioners.

To mitigate these problems, among others, SM ontologies have emerged over nearly three decades, along with the definitions of SM terms in related standards. A seminal ontological work was published by Kitchenham et al. in 1999 [2], followed by Ruiz et al. in 2004 [3], and Anquetil et al. in 2006 [4]. Regarding the definition of SM terms included in standards, the ISO/IEC/IEEE 14764 standard [5] was recently updated in 2022, which provides guidelines for software maintenance. This standard is based on the maintenance processes, activities, and tasks defined in ISO/IEC/IEEE 12207 [6], the latest edition of which was published in 2017. It is worth noting that [5] and [6] supersede the previous official documents from 2006 and 2008, respectively, and that ISO/IEC 14764 was first introduced in 1999, while ISO/IEC 12207 was introduced in 1995.

Next, we highlight some issues regarding the works cited above. On the one hand, just to exemplify a concept with some terminological heterogeneity (syntactic and semantic), let's look at the term Maintenance Request. This term label, which we have adopted in our proposal, is used only in Ruiz et al. In contrast, the term label Modification Request is used in Anquetil et al. and in ISO/IEC/IEEE 14764. Finally, Kitchenham et al. use the term label Maintenance Event, whose definition is also based on the semantics of an artifact. The terms Modification Request and Maintenance Request are often used interchangeably. However, they are not fully synonymous. A Maintenance Request is a type of Modification Request, but not all modification requests are maintenance requests. A Modification Request is a general term for any proposed change to a project's baseline in both the development and maintenance phases, while a Maintenance Request is a specific type of change that occurs in the ongoing support and upkeep of an already released software system and/or artifact.

On the other hand, existing SM ontologies fail to recognize the importance of representing concepts at different levels of generality and specificity. In particular, the authors do not consider a clear separation of concerns and the reuse of domain-independent and domain-dependent terms and relationships when conceiving generic/specific elements in their ontologies, nor do they consider their ontological components represented and situated within the context of a multilevel ontological architecture.

To bridge this gap, this paper proposes the Software Maintenance Top-Domain Ontology (SwMaintTDO), a high-level ontology developed within a multilevel ontological architecture. SwMaintTDO is primarily devoted to the domain-dependent concepts of a Maintenance Request registered by a Software Maintenance Organization and, consequently, to the necessary Managerial, Support, and Technical Processes and Activities, as well as the necessary Resources included in a Maintenance Project to modify the particular entities from the Current Situation to the Updated Situation supported by Maintenance Strategies, which can be Corrective, Adaptive, Preventive, or Perfective Maintenance Strategies depending on the specific purpose of a Maintenance Goal. These terms and relationships, among others, are located at the top-domain level in the context of a five-tier ontological architecture named FCD-OntoArch (*Foundational, Core, Domain, and instance Ontological Architecture for sciences*) [7].

The key contribution of this paper is the discussion of the SwMaintTDO concepts, relationships, and constraints, which align and harmonize with the domain-independent terms and relationships of ontologies at the core and foundational levels of FCD-OntoArch. In [8], the authors proposed some features that should be considered when developing ontologies. They indicate that the terms and relationships of the designed domain ontology must be based on a foundational ontology. In addition, they argue that the built ontology should be modular, extensible, and reuse already developed related ontologies, as indicated by the feature of being modular or integrated into a modular architecture. Other quality criteria that can be taken into account for the designed ontology are formal simplicity and transparency, also promoting the use of graphical representations for the conceptualization; completeness of coverage but, at the same time, conciseness and self-intuitiveness of the included elements; and balanced representation of both taxonomic (generic and partitive hierarchical) and non-taxonomic (associative) relationships [9, 10].

Considering only one of the above features, SwMaintTDO is based on terms and relationships of core ontologies for project management, situation, and process, which, in turn, their terms inherit the semantics of three concepts represented in ThingFO (*Thing Foundational Ontology*) [7]. ThingFO represents the world of any science with three very generic concepts, namely: the particular thing (as a concrete class or its instances), universal thing or categories of things (as a class of concrete classes), and human assertions that deal with these two concepts and their relationships. Having domain ontologies for different application fields that inherit elements of the same core and foundational knowledge base grants terms and relationships with similar semantics to have a shared semantic root.

As a result, the SwMaintTDO conceptualization has 77 main terms —each semantically enriched with terms reused from core ontologies or other top-domain ontologies—, 16 properties, 61 taxonomic and 67 defined non-taxonomic relationships, as well as UML constraints and 14 axioms specified in first-order logic. The functional requirements, representing the scope, are based on the formulation of 54 competency questions (CQs) that the aforementioned elements must answer to verify the coverage of the resulting conceptualization in relation to the CQs. Additionally, the ontology construction strategy used was Methontology [11].

It is worth noting that the proposed top-domain ontology provides a unified and extensible conceptualization that includes key and generic terms for maintenance processes, roles, resources, artifacts, goal purposes, and the transition of target modifiable and context entities from current to updated maintenance situations. Therefore, the generic but domain-dependent concepts of SwMaintTDO can be further specialized in low-domain ontologies that encompass more specific concepts, such as those that might be found in corrective or perfective maintenance ontologies, among others.

Finally, this paper illustrates the applicability and value of SwMaintTDO by using its terms and relationships with the vocabulary used in the process specification of a corrective maintenance strategy.

The remaining sections are structured as follows: Section II addresses related work, considering recognized SM ontologies and the terminology included in related standards. Section III analyzes the scope of SwMaintTDO, taking into account its place within an ontological architecture. It also analyzes the concepts, relationships, and constraints present in its conceptualization, as well as its axioms and verification aspects. Section IV demonstrates the applicability of the SwMaintTDO conceptualization by using its concepts and relationships in the process specification of maintenance strategies, particularly for corrective maintenance. Section V presents a discussion including some issues observed, and finally, Section VI draws the conclusions and future work.

## II. Related Work

In this section, we only analyze the recognized ontologies in the literature for the domain of SM, along with the SM terms in glossaries of related standards, excluding maintenance terminologies for other engineering disciplines.

Taking into account the features and quality criteria described above for ontologies, Kitchenham et al. [2], Ruiz et al. [3], and Anquetil et al. [4] use UML-based graphical representations for their ontological conceptualizations. They also represent sub-ontologies or modules within the main SM ontological component, thus considering the modularization criterion.

In [2], the following modules are represented: the Maintained product (sub)ontology, the Maintenance activity (sub)ontology, the Maintenance procedure (sub)ontology, the Maintenance organization process (sub)ontology, and the Peopleware (sub)ontology, which are integrated into the so-called Software Maintenance Ontology. In [3], the representation of the SM Project Ontology includes the following modules: the Maintenance Ontology —which, in turn, contains the sub-ontologies of Products, Activities, Processes, and Agents—, the Workflow Ontology, and the Measure Ontology. Finally, in [4], the ontological representation for SM includes five sub-ontologies, namely: Software System, Computer Science Skills, Modification Process, Organizational Structure, and Application Domain.

On the one hand, all three proposals use both taxonomic and non-taxonomic relations. If a proposal only used taxonomic relationships, it would be a taxonomy rather than an ontology. On the other hand, although the previous SM ontologies are modularized, they do not benefit from the semantic reuse of concepts provided by ontological components located at more generic levels in the framework of a multi-tier ontological architecture: they intermix domain-independent and domain-dependent terms and relations in the sub-ontologies that make up the SM ontology. In other words, having an SM domain ontology that inherits syntactic and semantic elements from the core and foundational knowledge base ensures that terms and relations with similar semantics have a shared semantic root, as we discuss later.

Regarding the definition of SM terms included in the standards, ISO/IEC/IEEE 14764 [5] includes 14 defined terms, often with notes, also indicating that other definitions given in ISO/IEC/IEEE 12207 [6] apply. Lastly, SWEBOK (v4.0a) [12] includes Chapter 7 dedicated to topics within the SM knowledge area, where the terminology used is primarily based on [5].

The glossaries of the cited standards are not comprehensive enough to form a mental representation of the concepts and relationships in the complex domain of SM. Furthermore, they lack graphical representations that would facilitate understanding and communication. ISO 704 [13] establishes the basic principles and methods for the preparation and production of terminologies, and in its previous edition stated: "*The more complex a concept system is, the more useful it is to clarify relations among concepts by representing them formally or graphically.*"

## III. Analysis of the Proposed Software Maintenance Ontology

This section is organized into four subsections. In III.A, we define the goal for the development of SwMaintTDO and its place within FCD-OntoArch. Subsection III.B formulates the scope in terms of categorized CQs. Subsequently, in section III.C, we analyze the resulting conceptualization of SwMaintTDO and present a subset of terms and relationships that will be used in section IV. Finally, in subsection III.D, we show an extract of the verification matrix of terms, relationships, properties, and axioms of SwMaintTDO considering CQs.

### *A. Location of the SwMaintTDO Component*

The goal statement for the development of SwMaintTDO can be summarized as follows: "To develop a top-domain SM ontology where its generic domain-dependent concepts related to maintenance processes, activities, roles, artifacts, and resources can be specialized at a low-domain level for various ontologies and applications and, as benefits, can be used as a common reference vocabulary for understanding, learning, and communication for stakeholders in both academia and industry, in addition to providing semantic processing capabilities."

Considering the above goal, the new SwMaintTDO component should be located at the top-domain level in the context of an ontological architecture. As a result, Fig. 1 shows its location within FCD-OntoArch.

SwMaintTDO is composed of domain-dependent terms and relationships for maintenance project-based processes, activities, roles, artifacts, outcomes, and resources, where a strategy is a type of resource. It extends and defines more specialized terms and relationships from core ontologies such as ProjectCO [14], ProcessCO [9], and SituationCO [7], and also reuses some terms from top-domain ontologies such as NFRsTDO (Non-Functional Requirements) [15], FRsTDO (Functional Requirements) [16], TestTDO [17], and MEvalTDO (Measurement and Evaluation) [18].

Note that Fig. 1 exhibits, with blue dashed lines, the dependencies of SwMaintTDO with other components located in the five-level ontological architecture. Also note that, on the left of this figure, the five levels of generality/specificity are shown, as well as the domain-independent and domain-dependent tiers. The components on the right correspond to the already developed ontologies, correctly aligned and harmonized.

For the sake of exemplification, in Fig. 2, the term Maintenance Request is stereotyped with the term Artifact — meaning that it has the semantics of an Artifact—, a term from ProcessCO. If we examine the ProcessCO ontology, we observe that the term Artifact is a type of Work Product, and this term is further stereotyped (semantically enriched) by the term Thing from ThingFO.

### *B. Stating the Scope in terms of Competency Questions*

The authors in [19] surveyed 63 ontology engineers, who concluded that CQs have proven useful in helping to define the scope (also known as functional requirements) of a given ontology, as well as in verifying its conceptualization.

As we did in previous proposals for ProjectCO, MEvalTDO, and TestTDO, we have designed semantic categories to which the main CQs can belong. The resulting 7 terminological categories also proved useful in identifying conceptual blocks or patterns and specifying the scope of SwMaintTDO in terms of 54 CQs, each of which belongs to only one category. As a result, 5 CQs belong to the SM Organizational Entity category, 3 to the SM Project Goal, 9 to the SM Situation, 12 to the SM Project Work Process, 10 to the SM Work Product, 9 to the SM Project Resource, and 6 CQs are included in the SM Role category. For example, below, CQ2 is included in the SM Organizational Entity, CQ9 and CQ13 are CQs related to the SM Situation, and CQ47 belongs to the SM Project Resource category.

- CQ2. Is there an organizational entity and function that serves as the single point of contact for receiving, recording, tracking, and communicating maintenance requests?
- CQ9. What are the types of maintenance situations that represent the states of entities before and after the application of maintenance project processes?
- CQ13. Under what circumstances are entities included in a maintenance situation considered to have the semantics of developable entities?
- CQ47. What are the types of a maintenance engineering strategy?

The rest of the CQs are found in [20] (pp. 35-37).

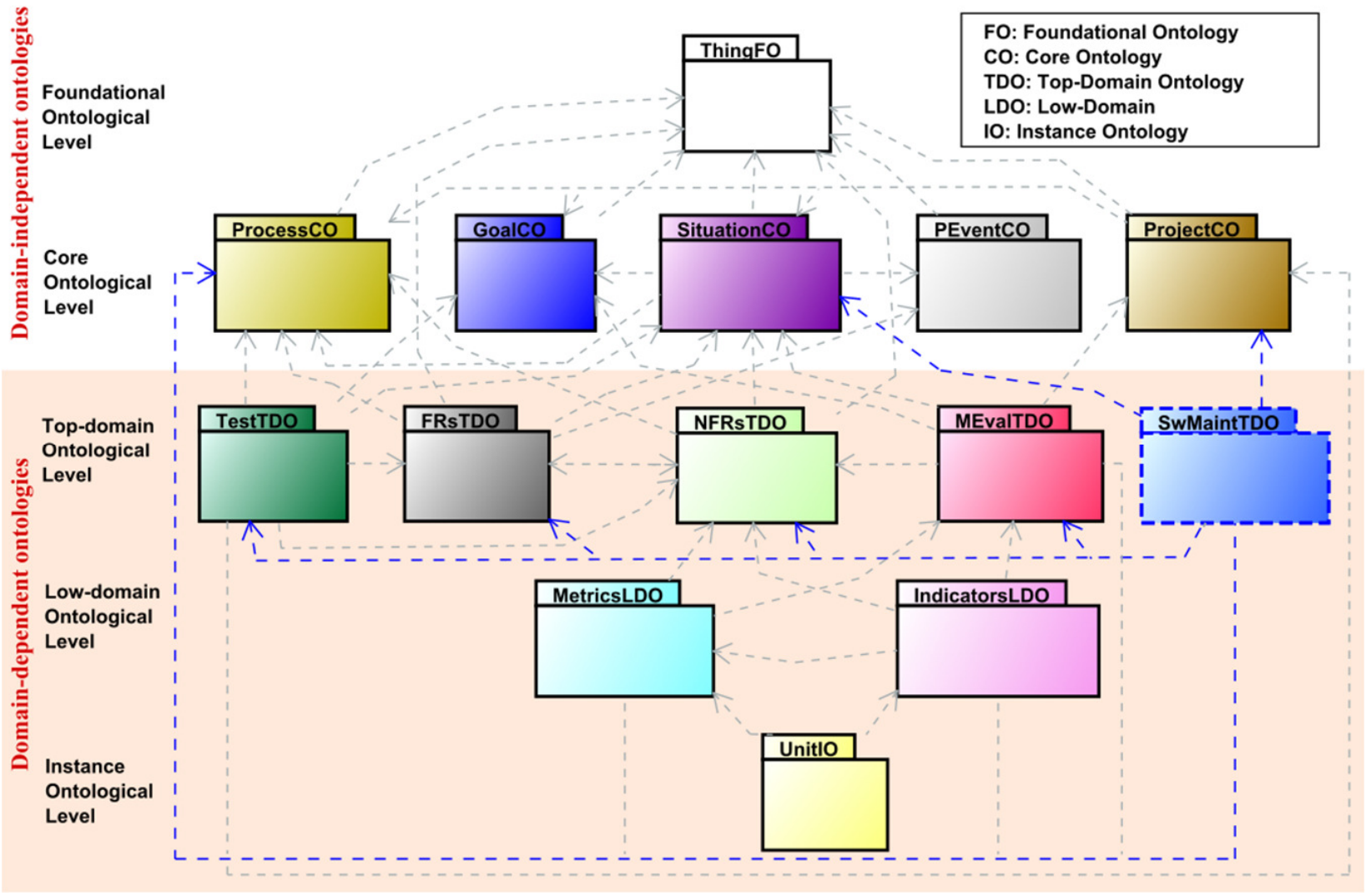

Fig. 1. Allocating the SwMaintTDO component or module in the context of the five-level ontological architecture, so-called FCD-OntoArch.

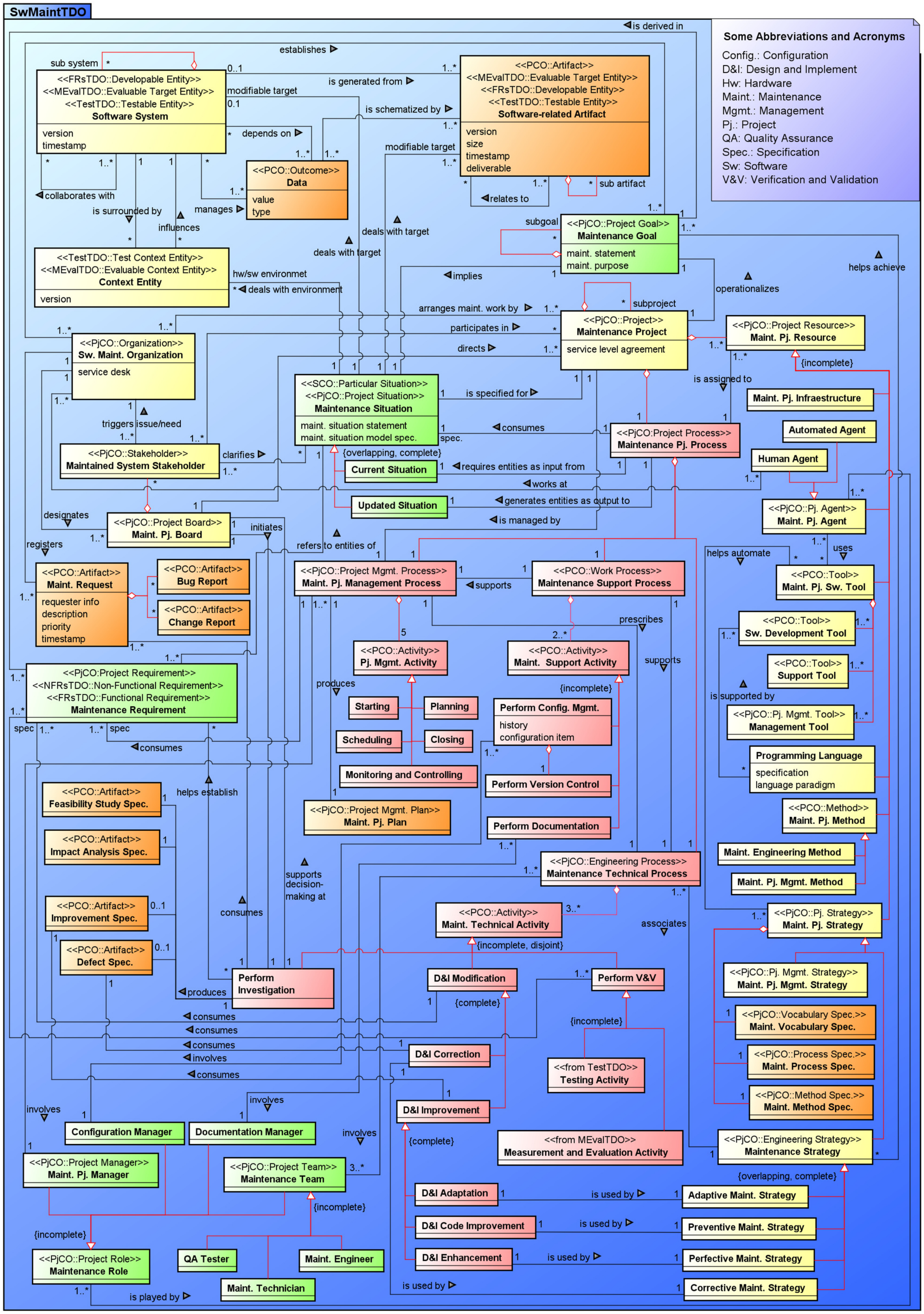

Fig. 2. SwMaintTDO v1.0: Top-Domain Ontology for Software Maintenance, which is located at the top-domain level of FCD-OntoArch in Fig. 1. Note that PjCO stands for Project management Core Ontology, SCO for Situation Core Ontology, PCO for Process Core Ontology, NFRsTDO for Non-Functional Requirements Top-Domain Ontology, FRsTDO for Functional Requirements Top-Domain Ontology, MEvalTDO for Measurement and Evaluation Top-Domain Ontology, and TestTDO for software Testing Top-Domain Ontology.

### C. SwMaintTDO's Concepts, Relationships, and Axioms

In the Introduction section, we have indicated the size of the elements included in SwMaintTDO. Most terms are semantically enriched with more generic terms reused from core ontologies and some others from top-domain ontologies. SwMaintTDO uses stereotypes as a mechanism to semantically enrich and align the terms, considering the reasons given in previous work, e.g., in [9].

Fig. 2 depicts the resulting conceptualization of SwMaintTDO specified in UML. In addition, some main terms are defined in Table I, but not all (77) are present, so it is recommended to consult their definitions in [20] (pp. 4-22). Next, we describe the elements of the SwMaintTDO conceptualization using the following text convention: top-

TABLE I. DEFINITION OF SOME TERMS OF THE SWMAINTTDO CONCEPTUALIZATION. ALL DEFINITIONS AND NOTES ARE AVAILABLE IN [20]

| SwMaintTDO Term | Definition |
|---|---|
| **Bug Report** | It is an Artifact, often part of a Maintenance Request, which formally documents a software defect or system failure to provide a clear and reproducible account of the issue supporting its resolution.<br>Note: A Bug Report typically includes a problem description, reproduction steps, the context in which it occurred, expected versus actual behavior, severity of the issue, and may also contain logs, screenshots, and other relevant information to support further investigation. |
| **Corrective Maintenance Strategy** | It is a Maintenance Strategy that specifies at least the Design and Implement Correction activity, including its sub-activities, tasks, and methods, necessary to reactively achieve the 'correct' purpose of a Maintenance Goal. |
| **Current Situation** | It is a Maintenance Situation that concerns the currently operating Target Entities —the Software System and its Software-related Artifacts— and, when applicable, the surrounding Context Entities, from which a Maintained System Stakeholder has raised an issue or need. |
| **Defect Specification** | It is an Artifact produced by the Perform Investigation Activity, based on the analysis of a Bug Report, that specifies the defect or flaw to be corrected, identifies the affected entities in the Current Situation, and defines the required modifications in the pertinent Software-related Artifacts to restore the Software System's expected behavior. |
| **Design and Implement Correction** | It is a Design and Implement Modification activity carried out to design and implement the necessary corrections to the Software-related Artifacts in the Current Situation, typically resulting in an updated Software System, based on the identified defects reported in the Defect Specification artifact. |
| **Feasibility Study Specification** | It is an Artifact produced by the Perform Investigation Activity that analyzes the feasibility of proposed solutions to a reported bug or proposed change, evaluating their technical, operational, and economic implications to support informed decision-making by the Maintenance Project Board. |
| **Impact Analysis Specification** | It is an Artifact produced by the Perform Investigation Activity that identifies and assesses the effects of a proposed improvement or defect correction on the Software System and its Software-related Artifacts in the Current Situation, including the scope of impact, affected entities, dependencies, and potential risks, to support Project Management Activities and decision-making by the Maintenance Project Board. |
| **Maintenance Goal** | It is a Project Goal for software maintenance that the Software Maintenance Organization intends to achieve through a specific Maintenance Project.<br>Note: Examples of maintenance (technical) purposes are: 'adapt', 'prevent', 'enhance', and 'correct'. |
| **Maintenance Process Specification** | It is a Process Specification that represents a model relating a set of Maintenance Project Process elements —such as Maintenance Activities, tasks, inputs and outputs, pre- and post-conditions, Maintenance Artifacts, and Maintenance Roles, among others— as a mandatory constituent of a Maintenance Project Strategy. |
| **Maintenance Project Board**<br>synonym: *Maintenance Steering Committee* | It is a formal, cross-functional steering committee operating within a Software Maintenance Organization, which is composed of key stakeholders of the maintained system, such as system owners, senior managers, and technical leads. Its primary function is to investigate, prioritize, and approve or reject all incoming Maintenance Requests related to a specific Software System and its Software-related Artifacts, serving as the central authority for all change control decisions. |
| **Maintenance Project Management Process** | It is a Project Management Process that constitutes one of the three sub-processes of the Maintenance Project Process, comprising a set of managerial activities intended to achieve the Maintenance Goal operationalized by a Maintenance Project. |
| **Maintenance Request** | It is an Artifact that formally describes and documents a Bug Report or a Change Report registered by a Software Maintenance Organization's service or help desk because a Maintained System Stakeholder raised an issue or need. |
| **Maintenance Requirement** | It is a Project Requirement, as well as a Functional Requirement and a Non-Functional Requirement, since it implies management constraints of the Maintenance Project in addition to engineering aspects of the entity requirements within the Maintenance Situation. |
| **Maintenance Strategy** | It is an Engineering Strategy implying four strategy types —Adaptive, Preventive, Perfective, and Corrective Maintenance Strategies— that specify the technical activities and methods necessary to achieve the various purposes of a Maintenance Goal. |
| **Maintenance Support Process** | It is a Work Process that constitutes one of the three sub-processes of the Maintenance Project Process, comprising a set of Maintenance Support Activities carried out to support configuration management, version control, and documentation of the entities in the Maintenance Situation. |
| **Maintenance Technical Process** | It is an Engineering Process that constitutes one of the three sub-processes of the Maintenance Project Process, comprising a set of Maintenance Technical Activities carried out to investigate the Maintenance Request and to perform the necessary modifications, verifications, and validations of the entities, resulting in the Updated Situation. |
| **Maintenance Vocabulary Specification** | It is a Vocabulary Specification that defines and structures the concept system specific to the maintenance area, domain, or discipline as a mandatory constituent of a Maintenance Project Strategy. |
| **Perform Investigation**<br>synonym: *Investigation Activity* | It is a Maintenance Technical Activity that consumes a Maintenance Request artifact and formally produces a set of documents, namely: Defect Specification, Improvement Specification, Impact Analysis Specification, and Feasibility Study Specification. |
| **Perform Verification and Validation**<br>acronym.: *Perform V&V* | It is a Maintenance Technical Activity implying two main activity types —Testing, and Measurement and Evaluation— carried out to verify and validate the performed changes (corrections, adaptations, enhancements, source code, or documentation improvements) to the entities, their components, or items in the Updated Situation. |
| **Updated Situation** | It is a Maintenance Situation that concerns the resulting deployed Target Entities —the Software System and its Software-related Artifacts— and, when applicable, the surrounding Context Entities, after the Maintenance Project Process and its sub-processes and activities have been completed and the parties have approved the established acceptance criteria for the modifications. |

domain terms begin with capital letters, properties are in italics, and relationships are underlined. In addition, it is worth noting that the colors used in some ontological elements of the diagram (Fig. 2) convey meanings to facilitate quick visual understanding. In the case of concepts, light green represents an Assertion term [7], while the other three colors represent a Thing (particular entity) term [7]: light red represents the semantics of a Process or Activity term; light orange represents the semantics of an Artifact or Outcome term; and light yellow represents the semantics of a Resource or an Organizational Entity, among other entity types. In the case of relationships, red conveys a taxonomic relationship, while black conveys a non-taxonomic relationship.

For the sake of brevity in the analysis, looking at Fig. 2, we can say that the Maintained System Stakeholder term represents the agent who can raise a problem or need (triggers issue/need) about the operated system to the Software Maintenance Organization. This concept includes individuals or groups, both internal and external to that organization, who have a vested interest in the ongoing operation and maintenance of a system. The *service* (or help) *desk* of the Software Maintenance Organization registers the Maintenance Request in the form of a Bug Report or Change Report document. Note that these last three terms have the semantics of Artifact, a concept defined by ProcessCO.

Taking into account a possible workflow to follow, the Maintenance Project Board —which governs the ongoing direction and evolution of the system by ensuring all maintenance activities are aligned with business priorities and the long-term health of the Software System— initiates the Investigation Activity, also known as Perform Investigation. The results of this activity (i.e., the Artifacts produced, such as the Feasibility Study Specification) support the decision-making of the board (also known as Maintenance Steering Committee), which ultimately makes an informed decision and authorizes or cancels the project's initiation (see the term Starting in Fig. 2, which is a subtype of the term Project Management Activity).

A Maintenance Project operationalizes a Maintenance Goal, a term that has two properties, namely *maintenance statement* and *maintenance purpose*. Instances of maintenance (technical) purposes are: 'adapt', 'prevent', 'enhance', and 'correct'. A Maintenance Goal implies a Maintenance Situation, of which, in turn, Current Situation and Updated Situation are the only subtypes. Table I defines these last two terms. Their definitions include the concepts of Software System and Software-related Artifact, which are key represented terms in Fig. 2, in the role of modifiable target entities within a given Maintenance Situation. The Software System typically includes deployed executables, services, APIs, and operational components that interact with Data and surrounding Context Entities within the operational environment. Whereas, typical Software-related Artifacts include the source code of software programs or web pages, Data schemas or templates, infrastructure code scripts, design diagrams, test case specifications, Maintenance Project Plans, among others. Note that, through appropriate transformations (e.g., compilation, deployment, or integration activities), Software-related Artifacts can become a Software System. In other words, a Software System is generated from one or more Software-related Artifacts.

On the one hand, the term (concept of) Maintenance Project mandatorily has a Maintenance Project Process, which in turn has three work processes that are labeled Maintenance Project Management Process, Maintenance Support Process, and Maintenance Technical Process. For example, the latter is semantically enriched by the stereotype <<Engineering Process>>, which is a term defined in ProjectCO. A Maintenance Project Process requires entities as input from a Current Situation and generates entities as output to an Updated Situation.

On the other hand, the term (concept of) Maintenance Project mandatorily has one or more Maintenance Project Resources, implying an interrelated set of resource types, such as Maintenance Project Infrastructure, Maintenance Project Agent, Maintenance Project Software Tool, Programming Language, Maintenance Project Method, and Maintenance Project Strategy, among other types defined in ProjectCO. Note that in the figure, a Maintenance Project Resource is assigned to a Maintenance Project Process.

Concerning engineering activities related to a Maintenance Technical Process, there are at least three types of Maintenance Technical Activities represented in Fig. 2, namely: Perform Investigation, Design and Implement Modification, and Perform Verification and Validation. For example, the term Design and Implement Modification is defined in [20] as "*It is a Maintenance Technical Activity implying two main activity types —Design and Implement Correction, and Design and Implement Improvement— carried out to produce the required changes (corrections, adaptations, enhancements, or source code improvements) to the Software System, its Software-related Artifacts, or a Context Entity in the Current Situation.*" Finally, it is worth noting that there are 14 axioms in [20] (pp. 31-34) that formalize aspects of some of the above terms and their relationships that are not explicit in the graphical conceptualization in Fig. 2. Thus, the description of axiom A11 says: "*If a Maintenance Project Process requires entities as input from a Current Situation, then its sub-processes (namely, the Maintenance Project Management Process, the Maintenance Support Process, and the Maintenance Technical Process) and its activities (namely, the Project Management Activity, the Maintenance Support Activity, and the Maintenance Technical Activity, respectively) requires these entities as well.*"

It is worth mentioning that CQ27 — What relationships exist between the sub-activities of designing and implementing modifications and the types of maintenance strategies? — requires that the links between technical activities and maintenance engineering strategies, as resources, be explicitly represented in the conceptualization. As a result, in Fig. 2, the reader can observe that the Design and Implement Correction activity is used by a Corrective Maintenance Strategy (both terms are defined in Table I); the Design and Implement Adaptation activity is used by an Adaptive Maintenance Strategy; the Design and Implement Code Improvement activity is used by a Preventive Maintenance Strategy; and the Design and Implement Enhancement activity is used by a Perfective Maintenance Strategy. These four strategies are subtypes of the term Maintenance Strategy enriched by the stereotype <<Engineering Strategy>>, which is a term defined in ProjectCO. In turn, the conceptualization shows a term labeled Maintenance Project Management Strategy, whose definition aims to represent the dynamics of management activities and methods of a Maintenance Project.

In short, any Maintenance Strategy can be considered a resource of a Maintenance Project that includes principles and integrated properties such as the Maintenance Process Specification, the Maintenance Method Specification, and the Maintenance Vocabulary Specification, to help achieve the *maintenance purpose* of a Maintenance Goal. To exemplify a strategy considering the behavioral and functional perspectives of process modeling [21], Section IV analyzes the Maintenance Process Specification only for the Corrective Maintenance Strategy.

Finally, regarding the modeled terms for the 6 CQs related to the SM Role category, Fig. 2 displays the 8 resulting main terms, as well as the taxonomic and non-taxonomic relationships. Note also that UML cardinalities act as constraints, as does the `{incomplete}` label for the subtypes of the term Maintenance Role. For example, a Maintenance Role is played by one or more Maintenance Project Agents. In addition, the Maintenance Support Activity called Perform Documentation involves a Documentation Manager role, which is a Maintenance Role.

### D. SwMaintTDO's Verification Matrix

Verification encompasses strategies that technically support the process of developing artifacts and systems, including principles, activities, methods, and tools to ensure quality during their development and maintenance. An ontology as a developed artifact can be conceptualized and/or implemented. In this work, we show the static verification of the SwMaintTDO conceptualization against the CQs. As a limitation, it should be noted that we did not focus on its implementation, which can strengthen, for example, verification based on SPARQL-based CQ testing on datasets.

As indicated in subsection III.B, the CQs represent the functional requirements related to the scope, from which the SwMaintTDO ontology was built, also considering the cited ontological architecture and related work. The primary purpose is to verify that all CQs were taken into account by the defined terms, properties, non-taxonomic and taxonomic relationships, constraints, and axioms.

As a result, Table II shows the verification matrix implemented using a black-box testing method. When designing this matrix, we considered as a test basis the 54 CQs and the elements that any conceptualized ontology has. The test checklist includes one row per CQ, in which we record what elements of the conceptualization help answer the CQ. After several iterations, at the end of the SwMaintTDO conceptualization process, we ensured that each CQ passed. Note that Table II shows an extract of the verification matrix. The complete verification matrix can be found in [20] (pp. 38-46).

In order to discuss some aspects, in the 2nd column of Table II, we see relationships with the underlined text and properties in italics. Thus, to cover CQ2, it is necessary to include five rows with the corresponding statements in order to answer it. For instance, the non-taxonomic relationship has the semantics of is used to express the dependency between a given term and the term that enriches it, as can be seen for the terms Software Maintenance Organization and Organization. Since the question in CQ2 begins with the phrase "…organizational entity…", we had to add the row indicating that the term Organization is a type of Organizational Entity, which is only represented in the ProjectCO component.

CQ13 requires being answered from the conceptualization, not only with the six rows with the corresponding statements, but also with two axioms (specified at the bottom of Table II) to formalize what is not explicit in the model. It is worth noting that the term Developable Entity is defined in FRsTDO as "*It is a Target Entity that represents a particular (concrete) object to be developed or modified, with a focus on its Features and Capabilities, serving as the primary point of interest within a particular situation.*" In turn, Feature is defined as "*It is a Property of a Developable Entity that refers to the intrinsic constitution, structure, or parts of it.*", and Capability as "*It is a Power of a Developable*

TABLE II. EXTRACT FROM THE CHECKLIST FOR INSPECTING HOW CQS ARE COVERED BY THE TERMS, PROPERTIES, RELATIONSHIPS, AND AXIOMS DEFINED BY SWMAINTTDO.

| **Competency Question (CQx)** | **What are the SwMaintTDO Terms, relationships, *properties*, Axioms, and their definitions/specifications that answer the corresponding CQ?** | **Actual Result (pass/fail)** |
|---|---|---|
| **CQ2.** Is there an organizational entity and function that serves as the single point of contact for receiving, recording, tracking, and communicating maintenance requests? | A Software Maintenance Organization has a *service desk* as a Property | pass |
| | A *service desk* is defined as "Organizational function within the Software Maintenance Organization that serves as […]" | |
| | A Software Maintenance Organization is a type of (has the semantics of) Organization | |
| | An Organization is a type of (has the semantics of) Organizational Entity (in ProjectCO) | |
| | A Software Maintenance Organization registers one or more Maintenance Requests | |
| **CQ13.** Under what circumstances are entities included in a maintenance situation considered to have the semantics of developable entities? | A Maintenance Situation deals with target one or more Software-related Artifacts | pass |
| | A Software-related Artifact is a type of (has the semantics of) Developable Entity | |
| | A Maintenance Situation deals with target zero or one Software System | |
| | A Software System is a type of (has the semantics of) Developable Entity | |
| | A Maintenance Requirement refers to entities of a Maintenance Situation | |
| | A Maintenance Requirement is a type of (has the semantics of) Functional Requirement | |
| | Related axiom: A3 (*) and A4 (**) | |

(*) A3 description/specification: *Any Software-related Artifact is a Developable Entity iff the Maintenance Requirement that refers to entities of the Maintenance Situation has the semantics of Functional Requirement.*

$$\forall sra, \exists mr, \exists ms: [Software_relatedArtifact(sra) \wedge hasTheSemanticsOfDevelopableEntity(sra) \leftrightarrow MaintenanceSituation(ms) \wedge MaintenanceRequirement(mr) \wedge hasTheSemanticsOfFunctionalRequirement(mr) \wedge refersToEntitiesOf(mr, ms) \wedge dealsWithTarget(ms, sra)]$$

(**) A4 description/specification: *Any Software System is a Developable Entity iff the Maintenance Requirement that refers to entities of the Maintenance Situation has the semantics of Functional Requirement.*

$$\forall ss, \exists mr, \exists ms: [SoftwareSystem(ss) \wedge hasTheSemanticsOfDevelopableEntity(ss) \leftrightarrow MaintenanceSituation(ms) \wedge MaintenanceRequirement(mr) \wedge hasTheSemanticsOfFunctionalRequirement(mr) \wedge refersToEntitiesOf(mr, ms) \wedge dealsWithTarget(ms, ss)]$$

*Entity that refers to what it does and behaves, or what it is supposed to do and behave.*"

Looking at Fig. 2, it can be seen that a Software System or a Software-related Artifact not only has the semantics of a Developable Entity, but also of an Evaluable and/or Testable Entity. Therefore, other axioms formalize these aspects.

## IV. Application of SwMaintTDO Concepts and Relationships in Maintenance Strategies

This section shows the applicability of the SwMaintTDO conceptualization as a common reference vocabulary, using its terms and relationships to model process and method specifications for a family of maintenance engineering strategies, as we did in [22] for measurement and evaluation strategies. For reasons of space, we only illustrate terminological reuse in the Maintenance Process Specification —not in the Maintenance Method Specification— for the Corrective Maintenance Strategy.

It is worth noting that most of the SwMaintTDO terms used in the Corrective Maintenance Strategy are defined in Table I. However, some others are specializations of this top-domain terminology and should therefore be represented at the low-domain level in FCD-OntoArch for a corrective maintenance ontology called CorMaintLDO, which is a work in progress.

Regarding the implication of the Maintenance Process Specification concept, a mandatory constituent part of a strategy, Fig. 3 depicts an activity diagram specified in the SPEM (*Software & Systems Process Engineering Metamodel*) notation, which represents functional and behavioral process perspectives (or views) that include maintenance processes, activities, and sub-activities, as well as their sequences and parallelisms, and input and output artifacts, among other SPEM elements.

Firstly, observing the parallelism, three coarse-grained processes are specified. At the top of the figure is the Maintenance Project Management Process, which should be broken down into sub-processes and activities for useful inclusion in a Maintenance Project Management Strategy specification. At the center of the figure is the Corrective Maintenance Process, a type of Maintenance Technical Process. This process is analyzed in detail below. And at the bottom of the figure is the Maintenance Support Process, which is also not broken down.

Secondly, when looking at the activities specified for the Corrective Maintenance Process, we see the activities called Perform Investigation (A1), Design and Implement Correction (A2), and Perform V&V (A3), which are terms reused from SwMaintTDO.

For A1, the first sub-activity is labeled Analyze Bug Report and Root Causes. This sub-activity consumes the Maintenance Request, specifically the Bug Report artifact, and produces the Defect Specification artifact. Note that a defect is a flaw or inconsistency in the source code, documentation, or Software System that causes incorrect, unexpected, or non-compliant behavior in some entities of the Current Situation, which must be corrected to restore the system to its intended functionality and compliance in the Updated Situation.

Regarding the reported bug, the diagram shows that it may be necessary for the Maintained System Stakeholder to provide additional information to analyze the issue. Davies and Roper [23] investigated how users report bugs in systems, considering what information they actually provide so that it can then help maintainers to identify and fix them. Their research demonstrates a clear discrepancy between the information that maintainers would expect to see in a Bug Report and the information that actually appears. Other subsequent Perform Investigation sub-activities shown in Fig. 3 are: Define Preliminary Corrective Solution, Perform Impact Analysis, and Perform Feasibility Study. For example, the second sub-activity mentioned produces the Impact Analysis Specification, and the third one produces the Feasibility Study Specification. These artifacts constitute relevant information for the Maintenance Project Board, which ultimately makes an informed decision and authorizes or cancels the project's initiation. Note that, for reasons of simplicity, the figure does not reflect or specify the interactions between technical and management processes.

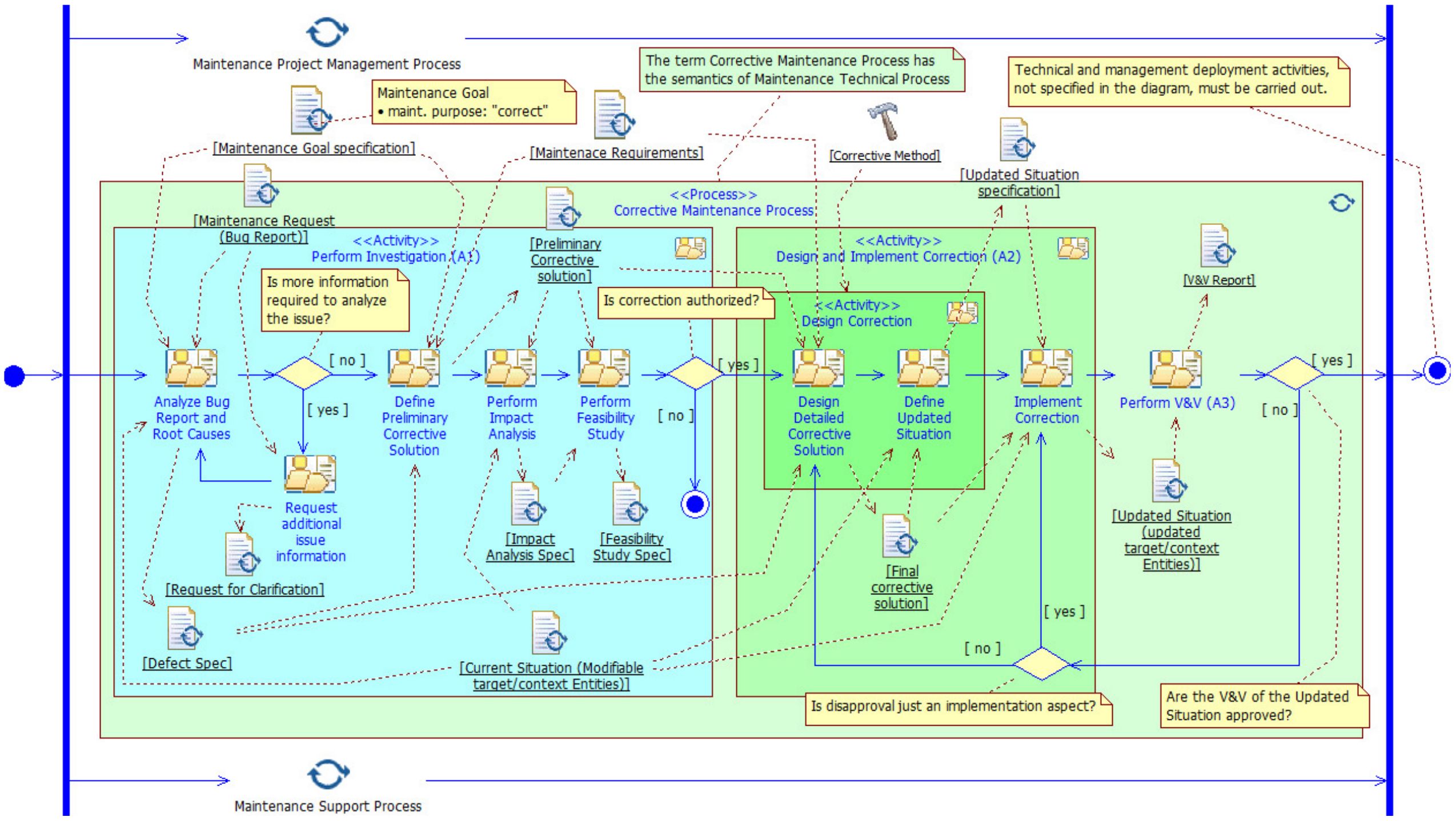

Fig. 3. Maintenance Process Specification for the Corrective Maintenance Strategy of the family considering the behavioral and functional perspectives.

Next, the A2 (Design and Implement Correction) activity is split down into Design Correction and Implement Correction sub-activities. In turn, the former is divided into Define Detailed Corrective Solution and Define Updated Situation.

It is worth noting that the SwMaintTDO terms referred to as Current Situation and Updated Situation in Fig. 2 are used in both A1 and A2, while Updated Situation is used in A3. For example, to carry out Analyze Bug Report and Root Causes, this sub-activity requires entities as input from a Current Situation —i.e., the current *version* of the Software System, its components, and the Software-related Artifacts, which are under configuration management and version control in a current operational context—, whereas the Implement Correction sub-activity generates entities as output to an Updated Situation.

Finally, regarding the engineering activities specified in the diagram in Fig. 3, activity A3 (Perform V&V) yields a V&V Report, which helps managers determine whether the entities in the Updated Situation are approved. If everything is correct, additional technical and management deployment activities, which are not shown in the diagram, must be carried out.

As a final consideration in this section, the SwMaintTDO conceptualization has proven useful as a common reference vocabulary not only for the process specification of the Corrective Maintenance Strategy illustrated above, but also for other maintenance engineering strategies in the family. For instance, considering the Maintenance Process Specification of the Perfective Maintenance Strategy, the names of the activities A1 and A3 are the same. Therefore, we observe commonalities within the family of strategies. However, the name of activity A2 changes to Design and Implement Enhancement, a term that appears in Fig. 2. Furthermore, for A1, some sub-activities are different, such as Analyze proposed Change and Define Maintenance Requirements. Others, such as Perform Impact Analysis and Perform Feasibility Study, remain unchanged.

Ultimately, thanks to the SwMaintTDO concept system —and other related low-domain level ontologies under development—, the specifications of the software Maintenance Strategies ensure terminological syntactic uniformity and semantic consistency, thus facilitating the understanding and communication of their processes, activities, tasks, methods, techniques, artifacts, and roles, among the different actors involved.

## V. DISCUSSION

Next, we analyze, primarily for the ontologies cited in Section II, certain heterogeneity and terminological discrepancies, as well as the lack of definitions for many terms related to maintenance engineering activities. Furthermore, we justify the selection of some terminological labels for these technical activities in SwMaintTDO.

Regarding the use of the term (label) Maintenance Activity —a synonym for Maintenance Technical Activity in SwMaintTDO [20]— it is only used in Kitchenham et al. [2] and Anquetil et al. [4], but its definition is not found in either ontology. In contrast, in ISO/IEC/IEEE 14764 [5], the defined term is called Software Maintenance, while in Ruiz et al. [3], it is simply called Maintenance without an explicit definition.

Considering the use of the term Modification Activity — a synonym for Design and Implement Modification in SwMaintTDO— it is only used in [2] and [4], which is defined in both cases. In contrast, in [3], the defined term is simply named Modification, while this term is not found in [5]. Note that in all three ontologies and in our proposal, Modification Activity is a type of Maintenance (Activity).

For example, Anquetil et al. define the term Modification Activity as "[it] *is a Maintenance Activity that may be Corrective Maintenance or Enhancement Maintenance (Adaptative, Preventive, or Perfective Maintenance).*" They graphically represent an inheritance relationship between the term Enhancement Maintenance and the terms Adaptive, Preventive, and Perfective Maintenance. However, none of these terms is explicitly defined.

Regarding the term Enhancement, it is also a type of Modification Activity in Kitchenham et al. The authors graphically represent an inheritance relationship between the terms Enhancement and New Requirements, Changed Implementation, and Changed Requirements. However, none of these terms is explicitly defined. In contrast, Ruiz et al. use the term Improvement instead of Enhancement (activity), although it is not defined. The term Improvement is a type of Modification. In turn, the following defined terms in [3], such as Preventing Problems (Preventive Maintenance), Changing Requirements (Perfective Maintenance), and Changing Implementation (Adaptive Maintenance), are subtypes of Improvement.

In summary, the label Improvement, at this level of the inheritance taxonomy, is only used in Ruiz et al. Conversely, in Kitchenham et al. (and in ISO/IEC/IEEE 14764), the label Enhancement with the semantics of activity is used, while in Anquetil et al., the defined term is called Enhancement Maintenance. Although often used interchangeably, Improvement and Enhancement have distinct meanings in Software Engineering. Improvement refers to modifying an entity element that is deficient or underperforming to correct weaknesses or inefficiencies, and occasionally implies augmenting something that already performs well. Enhancement, instead, refers to extending or enriching an entity that already functions adequately by adding new capabilities, features, or value. Accordingly, in SwMaintTDO, the Design and Implement Enhancement, Design and Implement Adaptation, and Design and Implement Code Improvement sub-activities are defined as subtypes of the Design and Implement Improvement activity. In turn, Design and Implement Improvement is a type of the Design and Implement Modification activity in Fig. 2.

The preceding discussion demonstrates the syntactic discrepancies in the naming of activity terms, as well as the lack of definitions for many of them, which often makes explicit semantic comparisons between existing terminologies difficult or impossible.

It is important to highlight that in SwMaintTDO, the labels Preventive Maintenance, Perfective Maintenance, Adaptive Maintenance, and Corrective Maintenance designate the concept of a strategy rather than an activity. As illustrated in Section IV, the term Maintenance Strategy has a broader semantics than that of a simple process or activity, encompassing not only process perspectives but also methods, among other aspects.

Finally, it should be noted that the term Maintenance Strategy, whether as an Engineering or Management project resource, is not found in the aforementioned vocabularies. Likewise, the concept of Maintenance Situation, with its subtypes Current Situation and Updated Situation, related to modifiable target and context entities, is also absent from these terminologies.

## VI. Conclusions

Software maintenance is a terminologically complex domain. Current conceptual discrepancies and variations continue to hinder understanding and effective communication between researchers and practitioners. To address this conceptual gap, we have presented the conceptualization and verification of SwMaintTDO, which takes into account relevant aspects of related previous work.

SwMaintTDO is conceived as a domain-dependent ontology for extensible generic terms and relationships for processes, activities, artifacts, roles, resources, goal purposes, and the transition of target modifiable and context entities from current to updated maintenance situations, oriented towards software maintenance projects. Consequently, it is a top-domain ontology developed in the context of a five-level ontological architecture. The layers of this ontological architecture promote the correct placement of components and alignment of concepts, fostering modularity and consistent extension and reuse of ontological elements at all lower levels.

Furthermore, we have analyzed not only the key concepts and relationships for the SM domain but also their applicability to enrich the vocabulary used in maintenance strategies. SwMaintTDO and related low-domain level ontologies that will be developed represent common reference terminologies with practical impact for different process and method specifications, as we have illustrated in Section IV. To date, in the related literature, there are no ontologies for this domain with the scope and coverage of our proposal.

As a work in progress, we are developing CorMaintLDO, the corrective maintenance ontology that contains more specific terms and relationships for corrective activities and methods, some of which are used in Fig. 3. In addition, we have carried out the Maintenance Process Specification of the Perfective Maintenance Strategy, but its corresponding low-domain ontology is still future work. The same applies to the maintenance engineering strategies dedicated to the goal's purposes of adapting and preventing.

## References

[1] D. L. Parnas, “Software aging”, 16th International Conference on Software Engineering, Sorrento, Italy, pp. 279–287, https://doi.org/10.1109/ICSE.1994.296790, 1994,

[2] B.A. Kitchenham, G.H. Travassos, A. von Mayrhauser, F. Niessink, N.F. Schneidewind, J. Singer, and H. Yang, “Towards an ontology of software maintenance”. Journal of Software Maintenance: Research and Practice, 11(6), pp. 365–389, 1999.

[3] F. Ruiz, A. Vizcaíno, M. Piattini, and F. García, “An Ontology for the Management of Software Maintenance Projects”, International Journal of Software Engineering and Knowledge Engineering, 14(03), pp. 323–349. https://doi.org/10.1142/s0218194004001646, 2004.

[4] N. Anquetil, K.M. de Oliveira, and M.G. Dias, “Software maintenance ontology”, In: Ontologies for software engineering and software technology. Springer Berlin Heidelberg. Calero, C., Ruiz, F., Piattini, M. (eds), Chapter 5, pp. 153–173. https://doi.org/10.1007/3-540-34518-3_5, 2006.

[5] ISO/IEC/IEEE 14764, “Software engineering —Software life cycle processes— Maintenance”, 2022.

[6] ISO/IEC/IEEE 12207, “Systems and Software engineering — Software life cycle processes”, 2017.

[7] L. Olsina, “The Foundational Ontology ThingFO: Architectural Aspects, Concepts, and Applicability”, In: Fred, A., et al. (eds.) Knowledge Discovery, Knowledge Engineering and Knowledge Management, IC3K 2021, Communications in Computer and Information Science, vol 1718, Springer, Chapter 5, pp. 73–99, 2023.

[8] M. D’Aquin and A. Gangemi, “Is there beauty in ontologies?”, Applied Ontology, 6:(3), pp. 165–175, 2011.

[9] P. Becker, M.F. Papa, G. Tebes, and L. Olsina, “Discussing the Applicability of a Process Core Ontology and Aspects of its Internal Quality”, Software Quality Journal, Springer, 30:(4), pp. 1003–1038. https://doi.org/10.1007/s11219-022-09592-3, 2022.

[10] L. Schneider, “How to Build a Foundational Ontology: The Object-Centered High-Level Reference Ontology”, A. Gunter et al. (Eds.): KI 2003, Springer, LNAI 2821, pp. 120–134, 2003.

[11] M. Fernández-López, A. Gómez-Pérez, and N. Juristo, “Methontology: From Ontological Art Towards Ontological Engineering”, Proc. of the Ontological Engineering AAAI-97 Spring Symposium Series, 1997.

[12] SWEBOK, “Guide to the Software Engineering Body of Knowledge v4.0a, Chapter 7, Software Maintenance”, H. Washizakipp (editor), IEEE Computer Society Products and Services, Sept 2025.

[13] ISO 704, “Terminology work – Principles and methods”, 2022.

[14] L. Olsina, P. Becker, and M.F. Papa, “The Project Management Ontology called ProjectCO: Architectural Aspects, Concepts, and Usefulness”, In: CIbSE 2024, Curitiba, Brazil, SBC_OpenLib, pp. 61–75, https://doi.org/10.5753/cibse.2024.28439, 2024.

[15] L. Olsina, P. Becker, M.F. Papa, and Rossi, G, “Ontological Concepts of Non-Functional Requirements Applied to Particular Situations in Measurement and Evaluation Projects”, CLEI Electronic Journal, vol. 28, issue 3, paper 7, http://dx.doi.org/10.19153/cleiej.28.3.7, 2025.

[16] L. Olsina and P. Becker, “FRsTDO v1.2's Terms, Properties, and Relationships -- A Top-Domain Functional Requirements Ontology”, Preprint in Research Gate, pp. 1–6, https://doi.org/10.13140/RG.2.2.11781.00486, 2025.

[17] P. Becker, G. Tebes, M.F. Papa, and Olsina L, “Enhancing a Software Testing Ontology”, ASSE'25, Argentine Symposium on Software Engineering, 11(2), 54th JAIIO, pp. 1–14, https://revistas.unlp.edu.ar/JAIIO/article/view/19521, 2025.

[18] L. Olsina, P. Becker, and M.F. Papa, “An Ontological Conceptualization of Measurement and Evaluation Activities and Methods and their Applicability in Evaluation Strategies”. 51st Latin American Computer Conference (CLEI'25), Valparaiso, Chile, IEEE Xplore, pp. 1–10, https://doi.org/10.1109/CLEI67442.2025.11420334, 2025.

[19] G.K.Q. Monfardini, J.S. Salamon, and M.P. Barcellos, “Use of Competency Questions in Ontology Engineering: A Survey”, In: Almeida, J.P.A., et al. (eds) Conceptual Modeling, ER 2023, LNCS, vol 14320, Springer, Cham., 2023.

[20] L. Olsina, P. Becker, and M.F. Papa, “SwMaintTDO v1.0's Concepts, and Relationships – A Top-Domain Ontology for Software Maintenance”, Preprint in Research Gate, pp. 1–48, https://doi.org/10.13140/RG.2.2.26543.19367, 2025.

[21] B. Curtis, M. Kellner, J. Over. “Process Modelling”, Communications of the ACM, 35(9), pp. 75–90, 1992.

[22] L. Olsina and P. Becker, “Family of Strategies for different Evaluation Purposes”, In: 20th Ibero-American Conference on Software Engineering (CIbSE’17) held in the framework of ICSE, CABA, Argentina, Published by Curran Associates 2017, pp. 221–234, 2017.

[23] S. Davies and M. Roper, “What's in a bug report?”, 8th ACM/IEEE International Symposium on Empirical Software Engineering and Measurement (ESEM'14), Torino, Italy, Article 26, pp.1–10. https://doi.org/10.1145/2652524.2652541, 2014.